\documentclass[apj]{emulateapj}
\usepackage{apjfonts}
\usepackage{graphicx}

\begin{document}

\title{FAR-ULTRAVIOLET OBSERVATIONS OF THE SPICA NEBULA AND THE INTERACTION ZONE}

\author{Yeon-Ju Choi\altaffilmark{1}, Kyoung-Wook Min\altaffilmark{1}
,  Kwang-Il Seon\altaffilmark{2}, Tae-Ho
Lim\altaffilmark{1},Young-Soo Jo\altaffilmark{1}, Jae-Woo
Park\altaffilmark{3}}

\email{email: zmzm83@kaist.ac.kr}

\altaffiltext{1}{Korea Advanced Institute of Science and Technology
(KAIST), 373-1 Guseong-dong, Yuseong-gu, Daejeon, Korea 305-701,
Republic of Korea}

\altaffiltext{2}{Korea Astronomy and Space Science Institute (KASI),
61-1 Hwaam-dong, Yuseong-gu, Daejeon, Korea 305-348, Republic of
Korea}

\altaffiltext{3}{Korea Intellectual Property Office (KIPO),
Government Complex Daejeon Building 4, 189 Cheongsa-ro,
Seo-gu,Daejeon, Korea 305-348, Republic of Korea}

\begin{abstract}

We report the analysis results of far ultraviolet (FUV)
observations, made for a broad region around $\alpha$ Vir (Spica)
including the interaction zone of Loop I and the Local Bubble. Whole
region was optically thin and a general correlation was seen between
the FUV continuum intensity and the dust extinction, except in the
neighborhood of the bright central star, indicating the dust
scattering nature of the FUV continuum. We performed Monte-Carlo
radiative transfer simulations to obtain the optical parameters
related to the dust scattering as well as the geometrical structure
of the region. The albedo and asymmetry factor were found to be
0.38$\pm$0.06 and 0.46$\pm$0.06, respectively, in good agreement
with the Milky Way dust grain models. The distance to and the
thickness of the interaction zone were estimated to be
70$^{+4}_{-8}$ pc and 40$^{+8}_{-10}$ pc, respectively. The diffuse
FUV continuum in the northern region above Spica was mostly the
result of scattering of the starlight from Spica, while that in the
southern region was mainly due to the background stars. The
\ion{C}{4} $\lambda\lambda$1548, 1551 emission was found throughout
the whole region, in contrast to the \ion{Si}{2}* $\lambda$1532
emission which was bright only within the \ion{H}{2} region. This
indicates that the \ion{C}{4} line arises mostly at the shell
boundaries of the bubbles, with a larger portion likely from the
Loop I than from the Local Bubble side, whereas the \ion{Si}{2}*
line is from the photoionized Spica nebula.
\end{abstract}


\section{INTRODUCTION}
Observations in the far ultraviolet (FUV) wavelengths (900-1750
{\AA}) provide a wealth of information regarding the physical and
chemical processes in the interstellar medium (ISM). For example,
many important ion emission lines associated with the cooling hot
gas of T $\sim$ $10^{4.5}$-$10^{5.5}$K, such as \ion{C}{3} $\lambda$
977, \ion{O}{6} $\lambda\lambda$ 1032, 1038, and \ion{C}{4}
$\lambda\lambda$ 1548, 1551, exist in this wavelength range
\citep{sla89,bor90,lan90}. The wavelength band also includes the
fluorescence emission lines of the Lyman transitions of molecular
hydrogen, which arise in the photodissociation regions (PDRs)
\citep{dop01,ste89}. Furthermore, the diffuse FUV continuum
background radiation probes the dust scattering of starlight
\citep{bow91,hen02,seo11a,seo11b}. Hence, a variety of astrophysical
targets have been analyzed based on FUV observations, such as
supernova remnants (SNRs; \citet{seo06,kim10a,kim10b}), superbubbles
\citep{kim07,jo11,jo12}, diffuse and molecular clouds, and
interstellar dust \citep{lee06,par09,psj12,lim13} .

FUV observations are also useful in the study of \ion{H}{2} regions
around bright stars. One example is the Spica Nebula. Spica
($\alpha$ Vir), located at (\textit{l}, \textit{b}) =
(316$^\circ$.11, 50$^\circ$.84), is a binary system with a primary
star of the B1 III-IV type and a secondary star of the B2 V type,
and is at the distance of $\sim$80 pc from the Sun in the direction
toward the Scorpius-Centaurus (Sco-Cen) association. The Spica
Nebula is relatively isolated, as revealed in the Wisconsin
H$\alpha$ Mapper (WHAM) image \citep{haf03}, and is characterized by
a low gas density of 0.2 -- 0.6 cm$^{-3}$ \citep{rey85,par10}.
\citet{par10} analyzed its spectral images made for the \ion{Si}{2}*
1533 and \ion{Al}{2} 1671 {\AA} lines and found enhancement of the
\ion{Si}{2}* intensity in the southern region, a feature also seen
in the H$\alpha$ image. They attributed the feature to the density
increase in the southern region. In contrast, the image of the
\ion{Al}{2} 1671 {\AA} line presented a broad central peak around
the central star without significant enhancement in the southern
region. The broad central peak was ascribed to the effect of
multiple resonant scattering. An ambient dust cloud was also
reported in the \ion{H}{2} region from the analysis of the Infrared
Astronomical Satellite (IRAS) 60 $\mu$m and 100 $\mu$m observations
\citep{zag98}. \citet{mur11} noted the FUV halo around Spica and
attributed it to dust scattering of the central star Spica. They
tried to constrain the distance to the dust layers with a single
scattering model. We will further discuss their results later for
comparisons with our model.

The Spica Nebula is located close to the so-called the "interaction
zone" of the Local Bubble (hereafter, LB) and the Loop I superbubble
\citep{egg95}. The existence of the LB was inferred from the soft
X-ray background observations \citep{cox87}. It is a low-density
($\sim$ $5.0\times10^{-3}$ cm$^{-3}$) region with a radius of $\sim$
100 pc, inside which the solar system is embedded. A number of
direct and indirect observations have revealed a cool and dense
neutral hydrogen gas layer with a column density
\textit{N}(\ion{H}{1}) of ~$\sim$ $1.0\times10^{19}$ cm$^{-2}$
\citep{cox87,lal03} as an expanding boundary of LB
\citep{knu78,fer85,fru94,wel94}. Nevertheless, the shape, origin,
and  the ionization structure of LB are still under debate
\citep{cox98,wel10,avi12}. The Loop I is a large radio loop centered
on (\textit{l}, \textit{b}) = (329$^\circ$$\pm$1$^\circ$.5,
17$^\circ$.5$\pm$3$^\circ$) with an angular radius of
$\sim$58$^\circ$. It is believed to have been formed by multiple
supernova explosions and/or strong stellar winds of the Sco-Cen OB
association located at $\sim$170 pc from the Sun. The boundary of
Loop I was observed to be expanding with a velocity of $\sim$20 km
s$^{-1}$ into the neutral ambient medium of density n $\sim$ 0.6
cm$^{-3}$ , to form a dense neutral shell with column density
\textit{N}(\ion{H}{1}) of $\sim$$1.0\times10^{20}$ cm$^{-2}$ at the
terminal shock \citep{sof74,egg95}. There is a wealth of evidence
that Loop I consists of a low density ($\sim$$2.5\times10^{-3}$
cm$^{-3}$), hot ($\sim$$10^{6.5}$ K) and highly ionized X-ray
emitting gas, generated by the interaction between the shock waves
from the recent supernova explosions within the Sco-Cen association
and the ambient neutral medium \citep{egg95,bre00,wil03,wel05}. The
X-rays, together with the young OB stars of the Sco-Cen association,
are believed to provide intense radiations that are responsible for
the ionization of neutral gas at the Loop I boundary, probably on
the inner side of the boundary wall \citep{wel05}.

The interaction zone is a huge ring-like feature of a dense neutral
matter within the apparent boundary of the Loop I shell, as
identified by \citet{egg95} from the analysis of diffuse X-ray and
21 cm neutral hydrogen maps. The feature has been suggested to be a
colliding structure of Loop I with LB. \citet{yos90} predicted the
ring-like feature using hydrodynamical simulations; when a shock at
the boundary of a bubble in a radiative stage makes contact with a
shock of another bubble, a dense neutral wall of a ring-like shape
with a density is 20 -- 30 times higher than those of the ambient
medium, would be formed at the intersecting points of the two
bubbles. \citet{egg95} noted that the \textit{N}(\ion{H}{1}) of this
region suddenly jumps from $\sim$ $10^{20}$ cm$^{-2}$ to
$\sim$$7.0\times10^{20}$ cm$^{-2}$ at a distance of $\sim$70 pc from
the Sun using the absorption measurements of metal ions by
\citet{fru94}, and they suggested the distance to the interaction
zone to be $\sim$70 pc. Assuming a toroidal shape of thickness of
$\sim$12$^\circ$, corresponding to $\sim$15 pc at 70 pc, its density
was estimated to be $\sim$15cm$^{-3}$. \citet{bre00} set the upper
limit of the distance of the interaction zone to be 80 -- 100 pc
since Loop I, being still active with ongoing star formations, may
have a higher pressure than the adjacent LB and push the interaction
shell towards the LB. \citet{cor04}, using the Str\"{o}mgren
photometry and \ion{Na}{1} column density measurements, estimated
the distance to the interaction zone to be 120 -- 150 pc. On the
other hand, \citet{rei08} suggested that the ring-shaped interaction
zone may be folded and warped as both the distance to the ring and
the color excess values vary along the ring; they estimated the
distance to the interaction zone to be 110$\pm$20 pc on the western
side and 280$\pm$50 pc on the eastern side.

In this paper, we present the analysis results of the FUV
observations of an extended region around the Spica Nebula that
includes the interaction zone. The main goal of this study is to
understand the morphological relationship between the Spica Nebula
and the neighboring interaction zone as well as the scattering
properties of the associated dust clouds using the FUV observations
and radiative transfer models.

\section{DATA}
 We employed two datasets for this study:
one is from the archival FUV data of Galaxy Evolution Explorer
(GALEX; \citet{mor07}), obtained as part of the All-Sky Imaging
Survey (AIS), and the other is the dataset used by
\citet{par07,par10}, obtained from the Far-ultraviolet Imaging
Spectrograph (FIMS; \citet{ede06a,ede06b}) aboard the Korean
microsatellite STSAT-1. Both observations cover very similar FUV
wavelength bands of 1350--1780 {\AA} and 1330--1720 {\AA} for GALEX
and FIMS, respectively. They are complementary to each other in that
the GALEX data have good statistics but no spectral information,
while the FIMS data provides spectral information but with rather
large statistical fluctuations due to relatively short exposure time
and lower sensitivity. Hence, we use the GALEX data for photometric
analyses as it gives more reliable estimations of the diffuse FUV
intensities than the FIMS data, and the FIMS data for spectral
analyses as it can discriminate the FUV line emissions from the
stellar continuum emissions. The GALEX sky background data (with an
extension of skybg), provided with a resolution of 16$'$
\citep{mor07} from the website
http://galex.stsci.edu/GR6/, were smoothed to obtain the final image
with a resolution of 0.5$^\circ$. The data reduction procedures of
the FIMS data are described in detail by \citet{par07}. More
information on the instrument FIMS can be found in
\citet{ede06a,ede06b}.


\section{RESULTS}

\begin{figure*}
 \begin{center}
  \includegraphics[width=5.1cm]{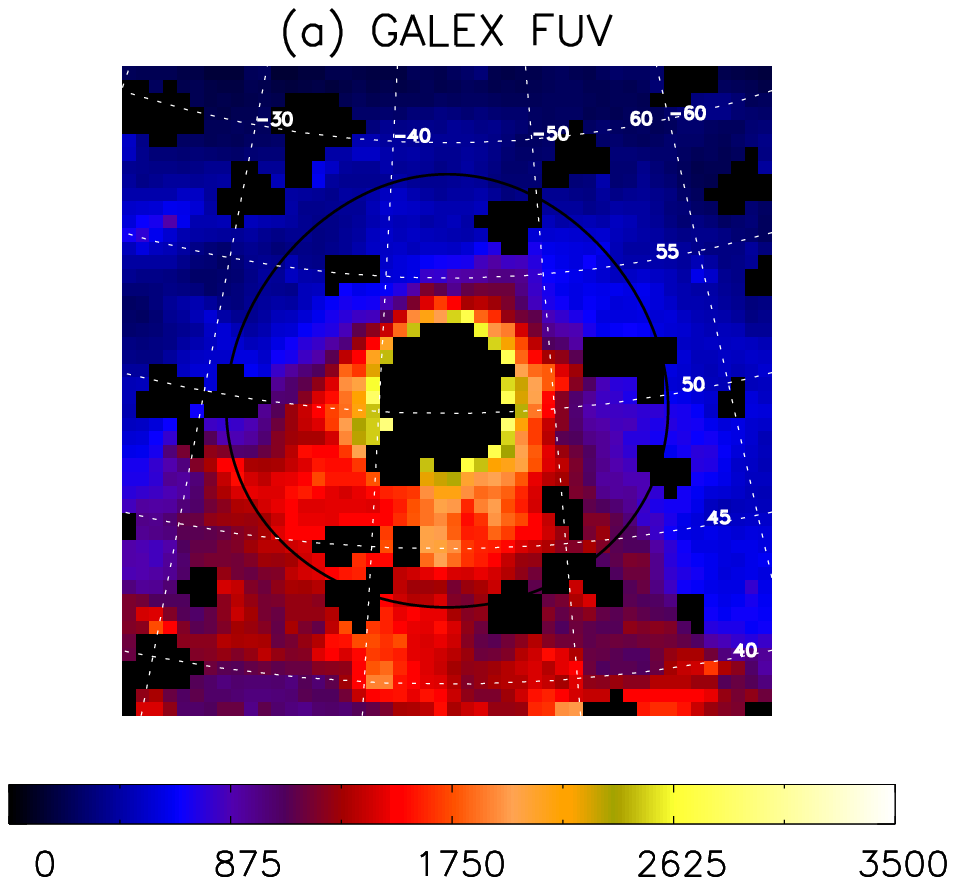}\hspace{0.8cm}
  \includegraphics[width=5.2cm]{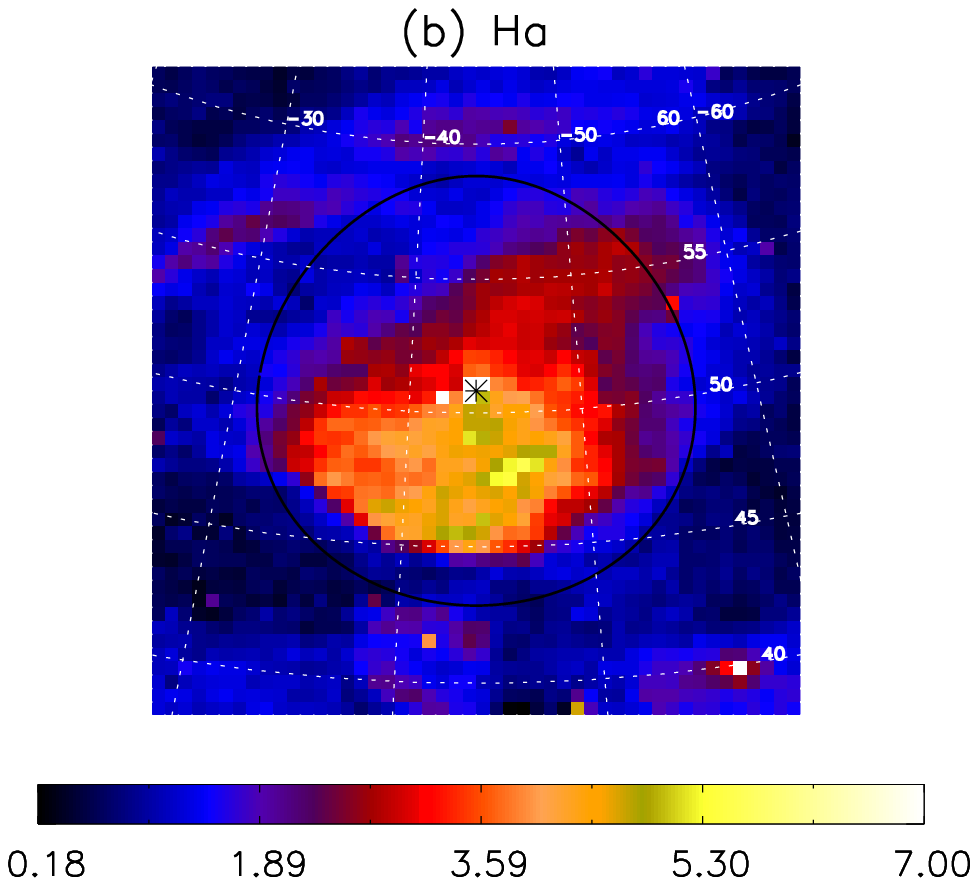}\\
  \vspace{0.8cm}
  \includegraphics[width=5.4cm]{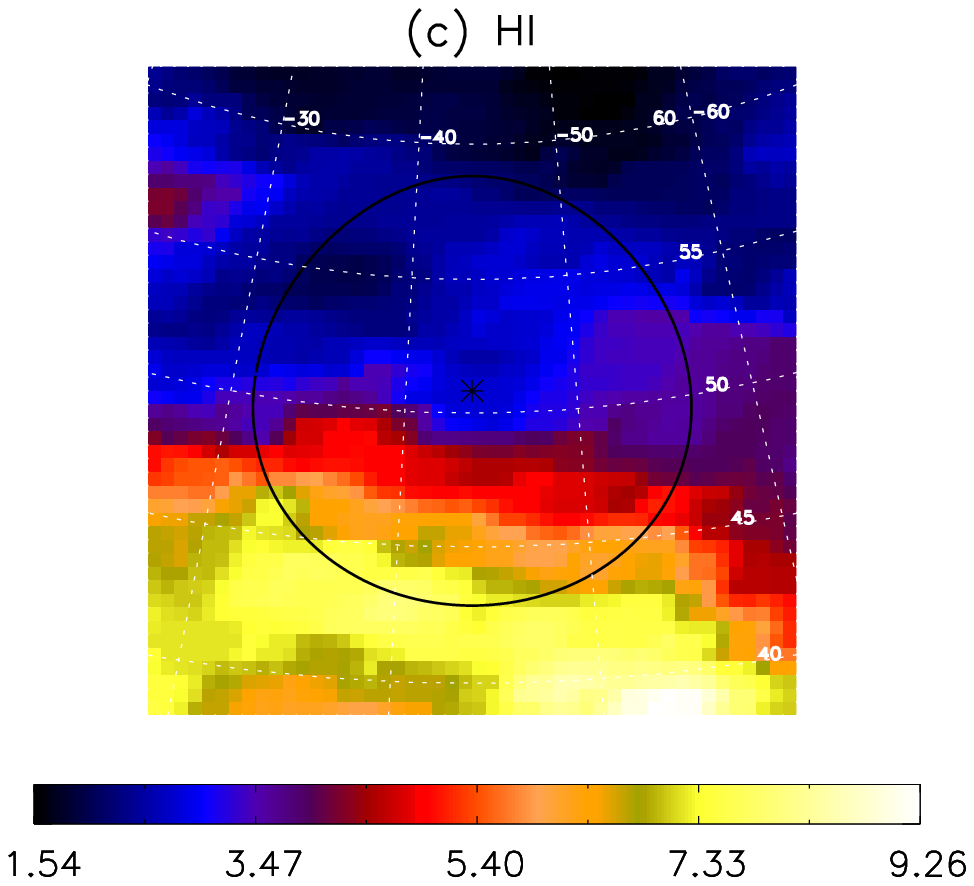}\hspace{0.5cm}
  \includegraphics[width=5.5cm]{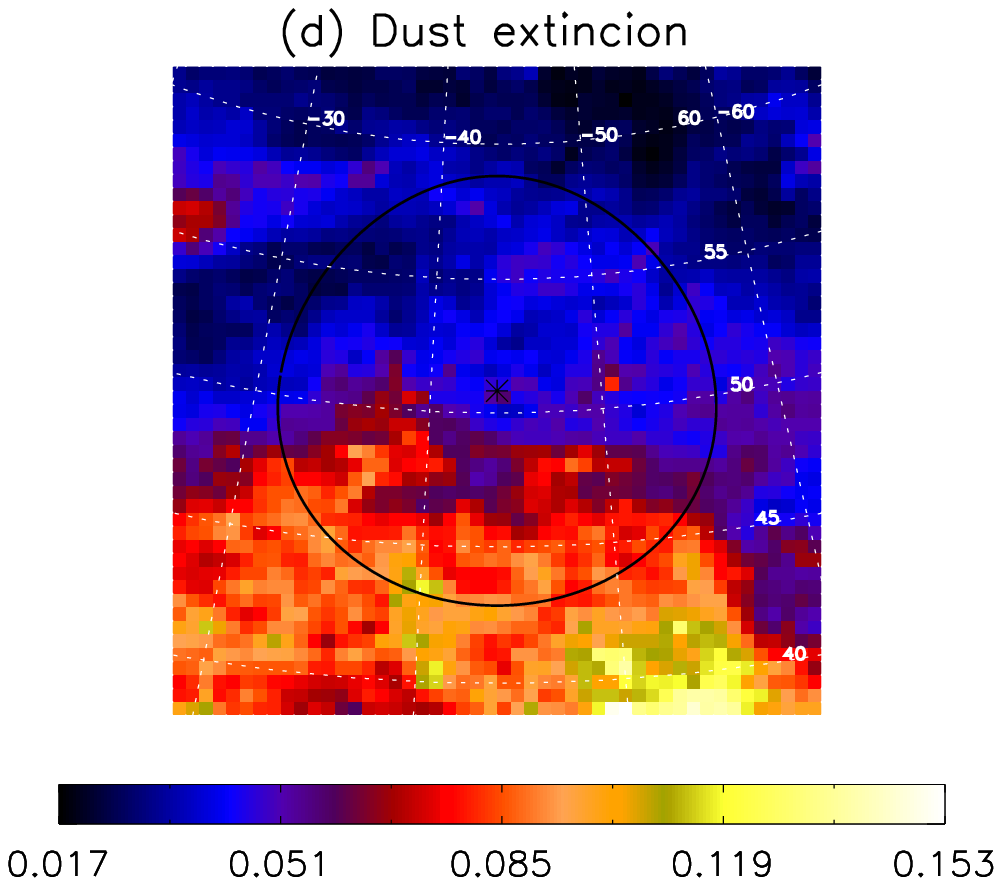}\\
 \end{center}

\caption{The extended Spica Nebula region observed in various
wavelengths: (a) the GALEX FUV intensity in CU ( photons sr$^{-1}$
s$^{-1}$ cm$^{-2}${\AA}$^{-1}$), (b) the H$\alpha$ intensity in
Rayleigh (R = (10$^{6}$/4$\pi$) photons cm$^{-2}$ sr$^{-1}$
s$^{-1}$), (c) the \ion{H}{1} column density in units of 10$^{20}$
cm$^{-2}$, and (d) the dust extinctioin given by
\textit{E}(\textit{B-V}). The images are shown in galactic
coordinates. The black circles of radius 8$^\circ$ denote the
boundary of the Spica Nebula. The central star $\alpha$ Vir is
marked with an asterisk at at (\textit{l}, \textit{b}) =
(316$^\circ$.11, 50$^\circ$.84).\label{fig:multi}}

\end{figure*}

Figure \ref{fig:multi} shows the GALEX FUV image, together with the
maps obtained from the observations made in other wavelengths. The
figures are shown for an extended region of 24$^\circ$ $\times$
24$^\circ$ around the Spica Nebula, in which the central star
$\alpha$ Vir is marked with an asterisk at (\textit{l}, \textit{b})
= (316$^\circ$.11, 50$^\circ$.84). The black circle of the 8$^\circ$
radius, defined as the boundary of the Spica Nebula, is also drawn
in each of the figures. In Figure \ref{fig:multi}(a), the missing
data regions shown in black are the regions of bright stars that
were intentionally avoided during the observations. The H$\alpha$
map of Figure \ref{fig:multi}(b), scaled in units of Rayleighs (1R =
(10$^{6}$/4$\pi$) photons cm$^{-2}$ sr$^{-1}$ s$^{-1}$), was
extracted from the H$\alpha$ sky survey map of \citet{fin03} and
\citet{haf03}. The \ion{H}{1} column density map of Figure
\ref{fig:multi}(c), shown in units of $10^{20}$ cm$^{-2}$, was taken
from the Leiden Argentine Bonn Galactic \ion{H}{1} Survey
\citep{kal05}. The dust extinction map of Figure \ref{fig:multi}(d)
was obtained from the Galactic reddening map of the \citet{sch98},
which was derived from the Cosmic Background Explorer (COBE) and the
Infrared Astronomical Satellite (IRAS) observations.

First, we note that the H$\alpha$ intensity is mostly confined
within a finite volume of the ionized nebula and drops abruptly at
the southern boundary near the interaction zone of LB and Loop I,
while the FUV intensity continues southward beyond the boundary of
the nebula. \citet{par10} argued that the \ion{H}{2} region is
"ionization bounded" at the boundary in the southern region with
\textit{N}(\ion{H}{1}) $\sim$ $6.0\times10^{20}$ cm$^{-2}$, below
which a good correlation is seen between the \ion{H}{1} column
density and the H$\alpha$ intensity. This conclusion accords well
with the fact that the estimated distance to the interaction zone is
$\sim$70 pc, which is comparable to that of the Spica Nebula.

There are two noticeable features in the FUV map: a halo around the
central star $\alpha$ Vir and an extended feature that generally
traces dust in the interaction zone. The FUV halo around $\alpha$
Vir is best seen in the northern region above the star and the halo
is a dust-scattered feature of the $\alpha$ Vir starlight. The
extended FUV feature in the southern region is predominant, as will
be discussed in Section 3, due to scattering of starlight
originating from other background stars by dust in the interaction
zone.

\begin{figure}
 \begin{center}
  \includegraphics[width=5.8cm]{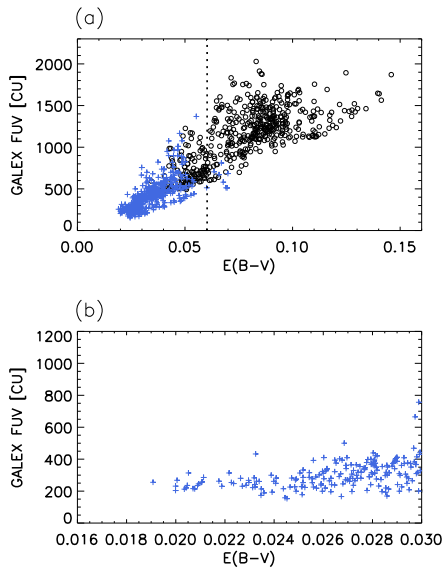}
  \end{center}
\caption{Correlation plots between the FUV intensity and dust color
excess in the region outside the bright circular region with an
angular radius of 3.5$^\circ$. (a) is for the entire region and (b)
is an enlarged plot for \textit{E}(\textit{B-V}) < 0.03. The data
points marked with blue crosses represent the northern region above
the central star $\alpha$ Vir and those with black circles represent
the southern region below the star. \label{fig:ebv}}
\end{figure}

Figure \ref{fig:ebv}(a) shows a pixel-to-pixel correlation plot
between the FUV intensity and the dust extinction in the whole
region of 24$^\circ$ $\times$ 24$^\circ$ except the bright central
part of an angular size of 3.5$^\circ$ around Spica. In the figure,
the data points with \textit{E}(\textit{B-V}) < 0.06, denoted in
blue, were mostly obtained in the northern region above the central
star, while those with \textit{E}(\textit{B-V}) > 0.06, denoted in
black, were obtained in the southern region. It should be noted that
the color excess \textit{E}(\textit{B-V}) in the whole region is
less than 0.14, which corresponds to the optical depth of $\tau$
$\sim$1 at 1565 {\AA}, and thus a general correlation between the
FUV intensity and dust extinction is expected if the radiation filed
is uniform \citep{hur94}. The figure does show a general trend of
increasing FUV intensity as \textit{E}(\textit{B-V}) increases,
except that the data points are severely scattered for the values of
\textit{E}(\textit{B-V}) from 0.04 to 0.08 while their intensities
are limited to $\sim$ 2000 CU. The severe scattering in the FUV
intensity is clearly a distance effect as the FUV intensity depends
strongly on the distance from the central star in this bright region
close to Spica, while the maximum value of $\sim$ 2000 CU is a
result of the brighter regions being excluded from the observations.
Nevertheless, excluding this large dispersion, we see an increasing
dispersion of the FUV intensity with increasing
\textit{E}(\textit{B-V}). \citet{seo11a} \& \citet{seo13} noted this
property in the analyses of the FUV continuum background, and
attributed it to the lognormality of the FUV intensity, caused by
the turbulence property of ISM. It is well known that the
probability distribution functions of the ISM density and column
density are very close to lognormal
\citep{vaz94,kle00,bur12,seo12a}. Since the dust-scattered intensity
is roughly proportional to the dust column density, the FUV
intensity should exhibit a signature of the lognormal density
structure. They also found that the dispersion of the FUV intensity
is consistent with that observed in the turbulent molecular clouds.

It is also interesting to see that, in Figure \ref{fig:ebv}(b),
which is an enlarged plot for the low extinction part of Figure
\ref{fig:ebv}(a), both the FUV intensity and the dust extinction
\textit{E}(\textit{B-V}) have minimum values of $\sim$200 CU and
$\sim$0.015, respectively. The minimum FUV intensity of $\sim$200 CU
is comparable to the so-called "isotropic component" of $\sim$300 CU
\citep{bow91,hen02,seo11a}, although some part of it may also come
from airglow \citep{suj10}. These minimum values could be regarded
as background values that are not relevant to the region considered
in this study.

\begin{figure}
 \begin{center}
  \includegraphics[width=4.3cm]{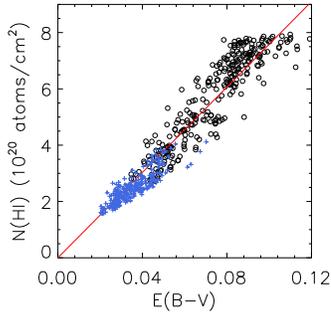}\\
  \end{center}
\caption{ Pixel-to-pixel comparison of \textit{N}(\ion{H}{1}) with
\textit{E}(\textit{B-V}), as obtained from Figure \ref{fig:multi}(d)
and Figure \ref{fig:multi}(c). The blue crosses and the black
circles are the data points obtained from the northern and the
southern regions, respectively. The red line represents the best-fit
ratio of \textit{N}(\ion{H}{1}) to
\textit{E}(\textit{B-V}).\label{fig:gasdust} }
\end{figure}

Figure \ref{fig:gasdust} shows a good correlation between the
neutral atomic hydrogen column density \textit{N}(\ion{H}{1}) and
\textit{E}(\textit{B-V}). The figure reveals that the ratio of
\textit{N}(\ion{H}{1}) to \textit{E}(\textit{B-V}) was
$\sim$$7.0\times10^{21}$ atoms cm$^{-2}$ mag$^{-1}$. There have been
several studies to estimate the ratio in the general diffuse
interstellar medium. Using 21 cm data, \citet{kna74} and
\citet{bur78} obtained an average ratio of $\sim$ $5.0\times10^{21}$
atoms cm$^{-2}$ mag$^{-1}$. \citet{sav79} used the Ly$\alpha$
absorption data obtained toward 100 stars by the Copernicus
spacecraft and found a ratio of $4.8\times10^{21}$ atoms cm$^{-2}$
mag$^{-1}$. \citet{dip94a} employed 393 sample stars observed with
the International Ultraviolet Explorer (IUE) satellite and obtained
$4.93\times10^{21}$ atoms cm$^{-2}$ mag$^{-1}$ in the regime of
\textit{E}(\textit{B-V}) $\leq$ 0.6, and $5.53\times10^{21}$ atoms
cm$^{-2}$ mag$^{-1}$ in the lower color excess values of
\textit{E}(\textit{B-V}) $\leq$ 0.3. The increase in the
\textit{N}(\ion{H}{1}) to \textit{E}(\textit{B-V}) ratio in the
lower color regime was attributed to the fact that a significant
amount of hydrogen is in a molecular form for
\textit{E}(\textit{B-V})
> 0.3. The present result is higher than these previous values,
implying that dust is probably less abundant in the shell boundaries
of the LB and Loop I. The evaporation of dust by the hot gas of LB
and Loop I and/or the expelling of dust by the strong radiations
pressure from the young OB stars of the Sco-Cen association might be
responsible for this deficiency of dust.

\begin{figure*}
 \begin{center}
   \includegraphics[width=5cm]{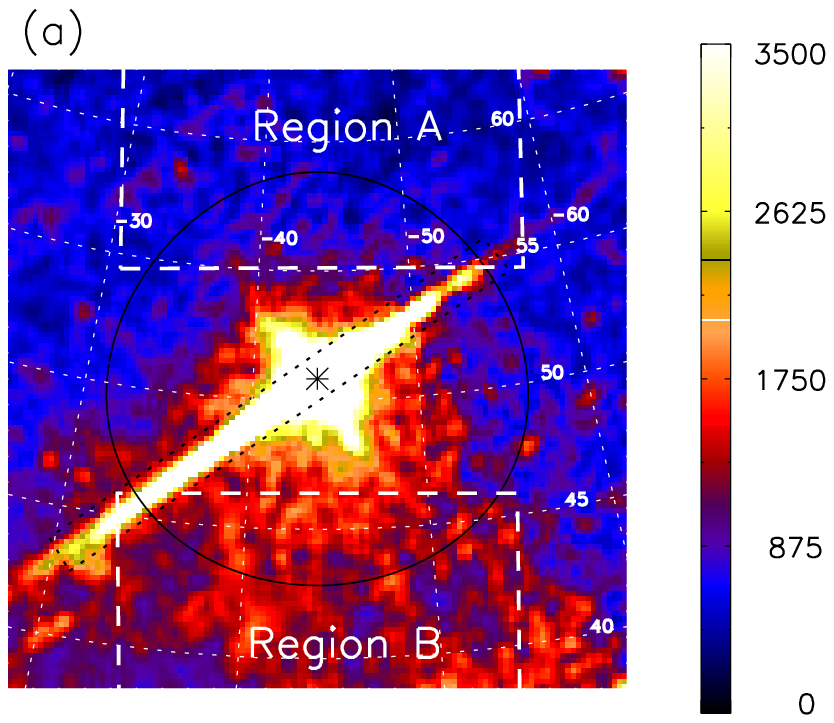}\\
  \vspace{0.5cm}
  \includegraphics[width=12cm]{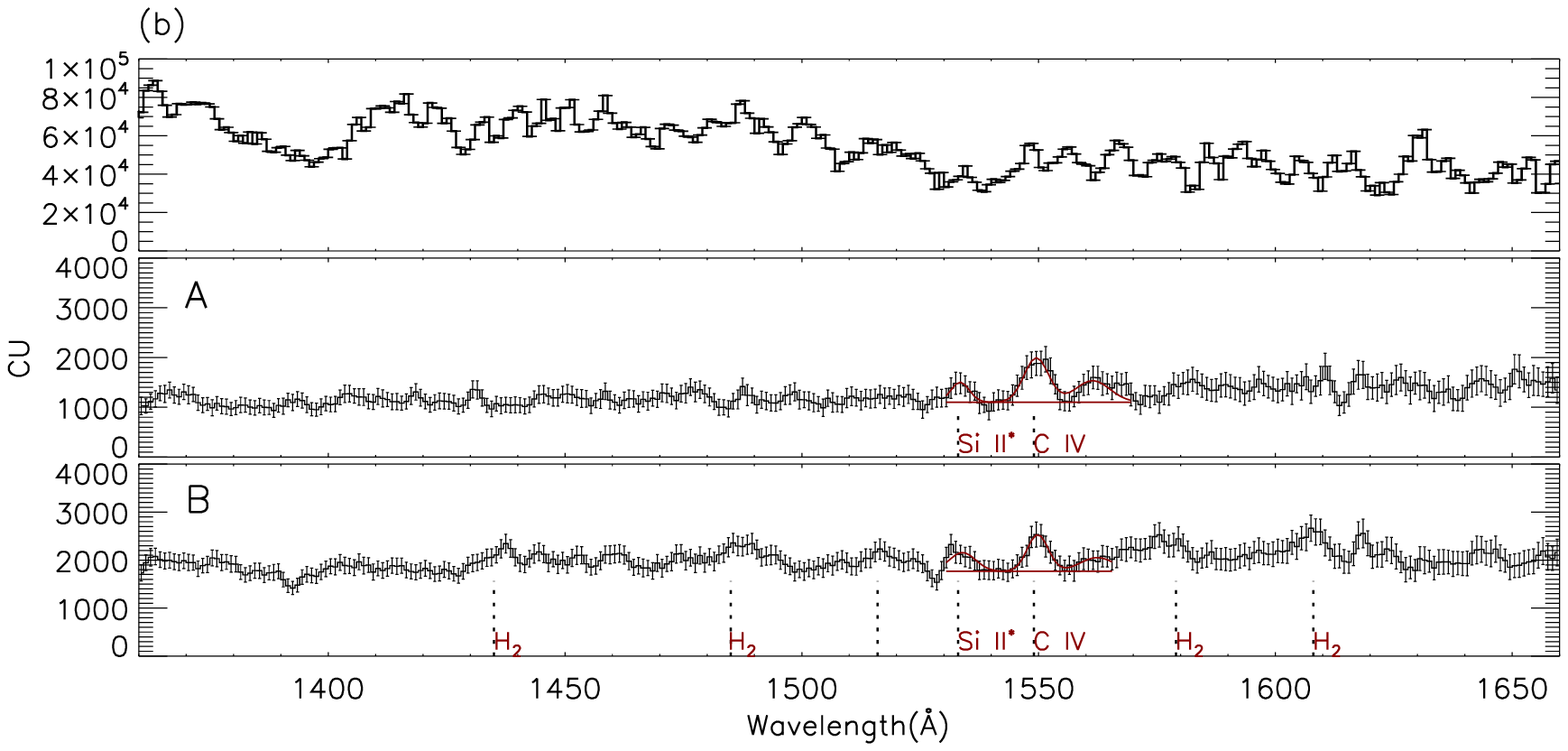}\\
  \end{center}
\caption{(a) The FUV intensity map observed by FIMS is given in CU (
photons sr$^{-1}$ s$^{-1}$ cm$^{-2}${\AA}$^{-1}$). The bright streak
in the narrow rectangular box is due to the instrumental scattering
of the bright central star along the slit direction. The black
circle of radius of 8$^\circ$ indicates the boundary defining the
Spica Nebula. The central star, $\alpha$ Vir, is marked with an
asterisk at at (\textit{l}, \textit{b}) = (316$^\circ$.11,
50$^\circ$.84). (b) Spectra observed by FIMS: the top panel shows
the spectrum for the central core region within an angular radius of
0.5$^\circ$ ; the middle and bottom panels present the spectra for
the regions A and B, as defined in (a), respectively. The vertical
dashed lines indicate the \ion{Si}{2}* and \ion{C}{4} lines, and the
brightest H$_2$ fluorescence lines. \label{fig:spectrum} }
\end{figure*}

The image of Figure \ref{fig:spectrum}(a), made by utilizing the
HEALPix scheme \citep{gor05} with a pixel resolution of 0.5$^\circ$,
was constructed from the FIMS data of the 1360 -- 1660 {\AA}
wavelength band, excluding the airglow line of $\lambda$1356 {\AA}
and the \ion{Al}{2} line of $\lambda$1671 {\AA}. We removed the
pixels of bright background stars and filled them by interpolating
the intensity from the neighboring pixels. A total of $\sim$50
bright pixels were removed based on the TD-$1$ stellar catalog. The
bright streak inside the narrow rectangular box is an artifact due
to the instrumental scattering of the bright central star along the
slit direction. The color levels are the same as those adopted for
the GALEX image in Figure \ref{fig:multi}(a). The two images of
GALEX and FIMS are very similar, although the FIMS intensity varied
less than the GALEX data because of smoothing of the FIMS image. In
Figure \ref{fig:spectrum}(b), the top panel shows the spectrum of
the core region inside an angular radius of 0.5$^\circ$, and the
middle and bottom panels show the spectra of the diffuse emission in
the two outer regions A and B, respectively. The spectrum of the top
panel is dominated by the central star ($\alpha$ Vir), and its shape
is consistent with the stellar spectrum of Spica observed by IUE.
The \ion{Si}{2}* and \ion{C}{4} ion lines are found in the outer
region spectra at wavelengths $\sim$1533 {\AA} and $\sim$1550 {\AA}
respectively. The \ion{Si}{2}* line was studied by \citet{par10}
using a photo-ionization model appropriate to the Spica Nebula. The
\ion{Si}{2}* emission was mostly observed within the Nebular and its
intensity was slighter higher in the southern part than in the
northern part. We fitted the \ion{C}{4} doublet lines observed in
the outer region spectra using the IDL-based MPFIT \citep{mar09} by
assuming two Gaussians functions, as done in \citet{seo06}; the
best-fit intensities are $\sim$ 5800 $\pm$ 1100 photons
s$^{-1}$cm$^{-2}$sr$^{-1}$(line unit; LU) and $\sim$4300 $\pm$ 1200
LU for the northern and southern regions, respectively. The large
errors in the results are mainly due to the fact that the results
were obtained from a wide region in which \ion{C}{4} intensity
fluctuated significantly and that the exposure time was rather
short. We also note that the molecular hydrogen fluorescence lines
are found in the spectrum of the southern region corresponding to
the interaction zone.


\section{MODELING AND DISCUSSION}

\begin{figure}
 \begin{center}
  \includegraphics[width=6.0cm]{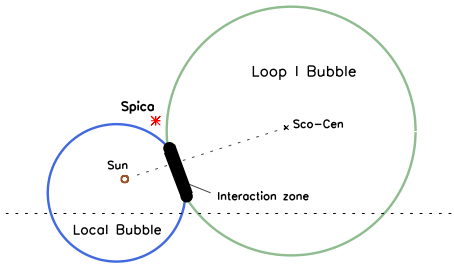}
  \end{center}
\caption{Schematic representation of the geometrical relation
between the Local Bubble, Loop I , the interaction zone and the
central star Spica. \label{fig:schematic}}
\end{figure}

Spica is likely located close to the interaction zone of the LB and
Loop I, as mentioned in the Introduction. Since Loop I is observed
to be expanding into the ambient medium of $\sim$0.6cm$^{-3}$, the
density of the ambient medium would be similar to that the Spica
Nebula region. \citet{rey85} and \citet{par10} estimated the density
of $\sim$0.6 cm$^{-3}$ and $\sim$0.22 -- 0.36 cm$^{-3}$
respectively, implying that Spica is located in the ambient medium
outside the two bubbles. It is also interesting to note that the
\ion{H}{1} column density to Spica was estimated to be
$\sim$10$^{19}$cm$^{-2}$ \citep{shu85,fru94}, which is similar to
the column density of the shell boundary of the LB
\citep{cox87,lal03}. The shell boundary of Loop I had a column
density of $\sim$10$^{20} $cm$^{-2}$ \citep{sof74,egg95}. These
observations suggest that Spica is located behind the LB shell but
in front of the Loop I shell. The density ($\sim$15 cm$^{-3}$;
\citet{egg95}) of the interaction zone was much higher than that of
the Spica Nebula. This indicates that Spica is located slightly away
from the interaction zone. Taking all these considerations into
account, the relative geometry between LB , Loop I and Spica could
be represented by a schematic diagram shown in Figure
\ref{fig:schematic}.

As we discussed previously, the interaction zone is rich in dust and
its column density \textit{N}(\ion{H}{1}) is proportional to the
level of dust extinction, implying that dust responsible for the
scattering of starlight should be located close to the central star
and the interaction zone. This will be confirmed by modeling the
scattered stellar photons by assuming dust distributions. A similar
study was conducted by \citet{mur11} for the halo around Spica
observed by GALEX. They focused on the region close to the star and
ignored the dust scattering effect in the interaction zone. They
employed a single dust-scattering model and assumed only a single
star (i.e, $\alpha$ Vir). In their first model, they placed a thin
dustsheet at a distance of 3 pc from the star, inferred from the 60
$\mu$m and 100 $\mu$m infrared observations, and obtained a very low
albedo value of 0.10$\pm$0.05 in the FUV band. In the second model,
they assumed an albedo of 0.4, and obtained a reasonable agreement
with the observation by employing two dust layers, one at 2 pc from
the star with 20\% of the total dust assigned to it and the other at
a distance of 30 pc from the star toward the Sun, with the remainder
of the dust allocated to it.

In our study, we present an improved model that includes the
dust-scattering effects of the interaction zone as well as the
radiation field originating from the background stars near Spica, in
order to study the extended region of the Spica Nebula. We employed
a three-dimensional Monte Carlo radiative transfer code, in which
photons are allowed to be multiply scattered (e.g.,
\citet{seo12b,jo12,lim13}). However, single dust-scattering was
dominant because the region is optically thin and the optical depth
approaches $\sim$1 in the interaction zone. The photon direction
after each scattering was modeled by the Henyey-Greenstein phase
function \citep{hen41}

\begin{equation}
    \phi(\theta) = \frac{a}{4\pi}\frac{(1-g^2)}{[1+g^2-2g\cos(\theta)]^{1.5}}
\end{equation}

where \textit{a} and \textit{g} are the albedo and the phase
function asymmetry factor of dust scattering, respectively. The
simulation code employed the peeling-off method \citep{yus84} for
simulation efficiency; while each photon from radiation sources
propagates through the dust medium and experiences multiple
scatterings into random directions, the code stores the scattered
fraction of the photon in the direction toward the Sun from each
scattering, and these fractions are added up to give the final
intensity. The parameters to be determined by the simulation are the
albedo a, the asymmetry factor g, and the spatial distribution of
dust. More detail descriptions on the simulation technique can be
found in Jo et al. (2012) and Lim et al. (2013).

The diffuse interstellar FUV emissions are composed of three
different components in addition to the dust-scattered starlight:
ionic emission lines from hot gases of 10$^{4.5}$ -- 10$^{5.5}$ K,
H$_2$ fluorescent emission lines originating from photodissociation
regions irradiated by interstellar UV photons, and two-photon
continuum emission from the ionized gas of about 10$^{4}$K.
Therefore, in order to obtain only the dust-scattering component
from the total observed FUV intensity, it is necessary to estimate
various contributions to the observed FUV intensity. In addition,
the isotropic background derived from the FUV intensity extrapolated
to \textit{N}(\ion{H}{1}) = 0 should be removed from the observed
FUV intensity \citep{bow91,hen02}. In this study, we subtracted the
minimum intensity of 200 CU as the contribution of the isotropic
background. The two-photon emission was estimated using a simple
formula of 60 CU per 1 Rayleigh of H$\alpha$ intensity
\citep{rey90}. However, the contribution is insignificant even in
the Spica Nebula with $\sim$6$\%$ contribution of the total FUV
intensity. The contribution of two-photon continuum emission in the
diffuse ionized gas or warm ionized medium outside of the Spica
Nebula is also unimportant (see \citet {seo11a}). The contribution
of H$_2$ fluorescent emission is also negligible as it was estimated
to be smaller than 10\% even for the interaction zone.

The simulation domain consisted of 400 $\times$ 400 $\times$ 400
rectangular grids cells each with cell size of 1 pc. The central
axis was chosen along the line of sight toward Spica, and the Sun
was placed at the center of the front face of the simulation box.  A
total of 21,225 stars from the TD-1 and Hipparcos catalogs were
disposed in the simulation box as radiation sources. A total of
$10^{8}$ photons were generated and assigned to the stars in
proportion to their intrinsic luminosities. The spatial distribution
of dust was obtained based on the Schlegel, Finkbeiner and Davis
(SFD) dust map \citep{sch98}. As the map is two-dimensional without
the distance information along the line of sight, a dust slab with a
thickness (t) was assumed to be located at a distance (d). The total
dust column density along each sightline was uniformly distributed
among the dust cells in the sightline. The optical depth at 1565
{\AA} was derived from the \textit{E}(\textit{B-V}) value of the SFD
map with \textit{R}$_V$ = 3.1, the average value for the Milky Way.
We assumed that the minimum \textit{E}(\textit{B-V}) of 0.015 found
in Figure \ref{fig:ebv} is not related to the dust around Spica and
subtracted the value from the dust extinction map. Comparison
between the simulation and the observation was made only for the
region with an angular distance larger than 3.5$^\circ$ from the
central star as the GALEX data may not be reliable for the bright
region close to the central star. The simulation parameters were
varied as follows: the distance to the front face of the dust slab
was varied from 50 to 90 pc in steps of 2 pc and the thickness from
10 pc to 50 pc in steps of 2 pc. We also varied the albedo (a) from
0.30 to 0.50 in steps of 0.02 and the asymmetry factor (g) from 0.40
to 0.60 in steps of 0.02. Each simulation was compared with the
GALEX observation to obtain a set of best-fit parameters.

\begin{figure}
 \begin{center}
   \includegraphics[width=3.5cm]{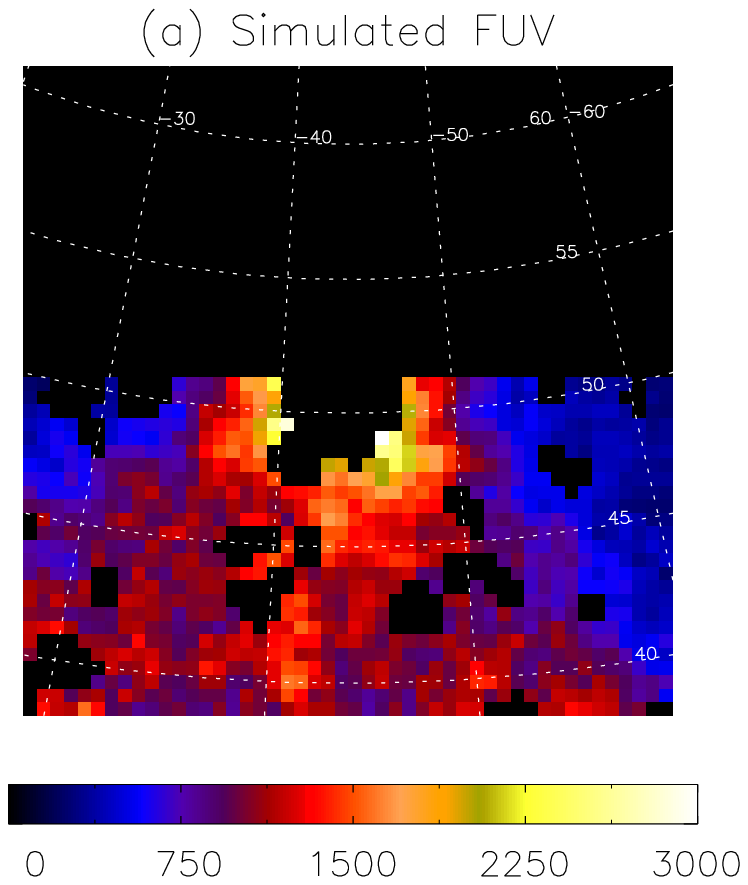}\hspace{0.4cm}
  \includegraphics[width=3.5cm]{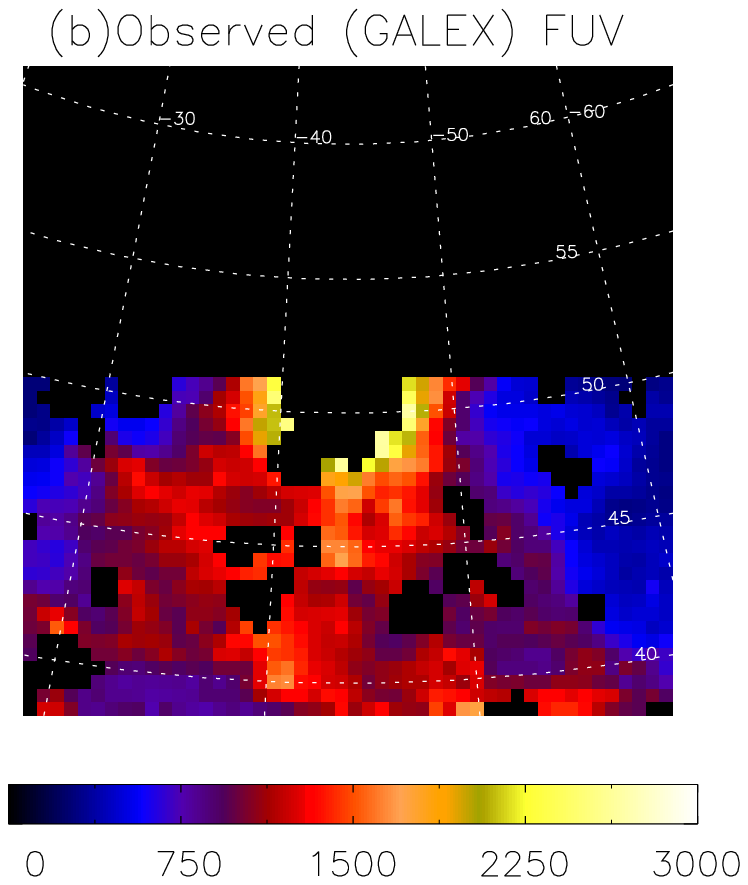}\\
  \vspace{0.8cm}
  \includegraphics[width=4.0cm]{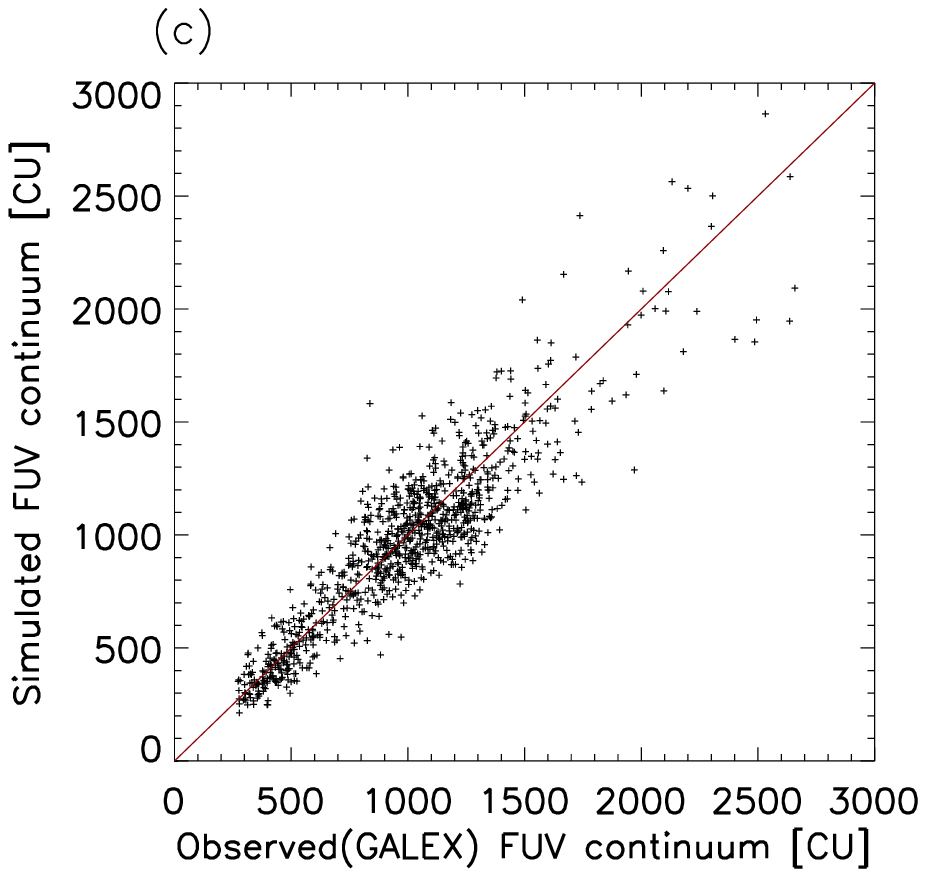}\\
  \end{center}
\caption{FUV continuum map of the southern region obtained from (a)
the dust scattering simulation, and (b) the GALEX observation. A
pixel-to-pixel correlation plot of the FUV intensity obtained from
(a) and (b) is shown in (c).\label{fig:south}}
\end{figure}

A Monte Carlo simulation of dust scattering does not constrain the
scattering parameters effectively when only a single star is
present, as discussed in \citet{mur11}. In fact, such a simulation
may produce many parameter sets that yield similar FUV intensities.
Hence, we first fitted the southern region to obtain the optical
parameters of dust, excluding the northern region where Spica
dominates the FUV intensity to reduce the effect caused by the
bright central star. The dust parameters determined from the
southern region will be adopted for the simulation of the northern
region. The resulting best-fit parameters were d = 70$^{+4}_{-8}$
pc, t = 40$^{+8}_{-10}$ pc, \textit{a} = 0.38$\pm$0.06 and
\textit{g} = 0.46$\pm$0.06. Figure \ref{fig:south} presents the
best-fit simulation map together with the GALEX map and compares the
calculated intensities with the observed value. The figure shows
good agreement between the simulation and the observation. It should
be also noted that \citet{egg95} suggested that the location of the
interaction zone was $\sim$70 pc, which agrees with the present
best-fit distance. Furthermore, the obtained optical parameters
\textit{a} = 0.38 and \textit{g} = 0.46 accord well with the studies
conducted for other objects \citep{jo12,lim13}, as well as the
theoretical predictions (\textit{a} = 0.4 -- 0.6 and \textit{g} =
0.55 -- 0.65) from the carbonaceous-silicate grains model
\citep{wei01}. Hence, we use a =0.38 and g = 0.46 for further
discussions.

\begin{figure}
 \begin{center}
   \includegraphics[width=5.5cm]{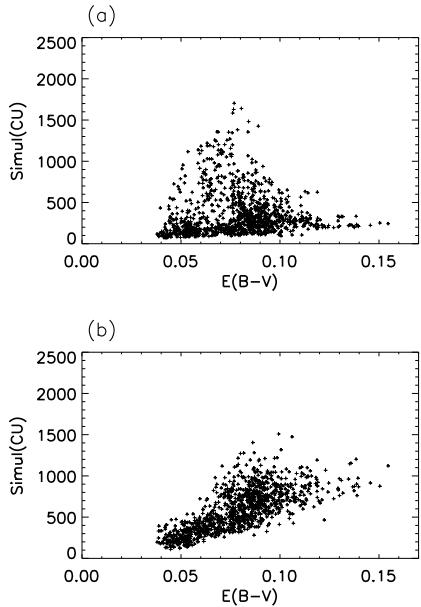}
  \end{center}
\caption{Simulated FUV intensity plotted against
\textit{E}(\textit{B-V}) for the southern region: (a) simulation
with the central star only and (b) simulation with the background
stars only.\label{fig:source}}
\end{figure}

The contributions of the central star and background stars to the
total FUV intensity are shown in Figures \ref{fig:source}(a) and
\ref{fig:source}(b), respectively. Figure \ref{fig:source}(a) shows
that the contribution from the central star is most important for
\textit{E}(\textit{B-V}) $\leq$ 0.06, which is not surprising, as
the distance from the central star increases with an increasing
\textit{E}(\textit{B-V}). On the other hand, the background stars
become more important than the central star for
\textit{E}(\textit{B-V})
> 0.06. In fact, the structures of the interaction zone in the
observed FUV map appear to match the simulation map obtained only
with the background stars, which implies that the observed FUV
intensity in the interaction zone is dominated by scattering of the
photons of the background stars. It is also interesting to note that
the simulated intensity is also limited by a maximum intensity of
1500 CU, due to the finiteness of the local stellar radiation field
in the region.

\begin{figure}
 \begin{center}
   \includegraphics[width=3.5cm]{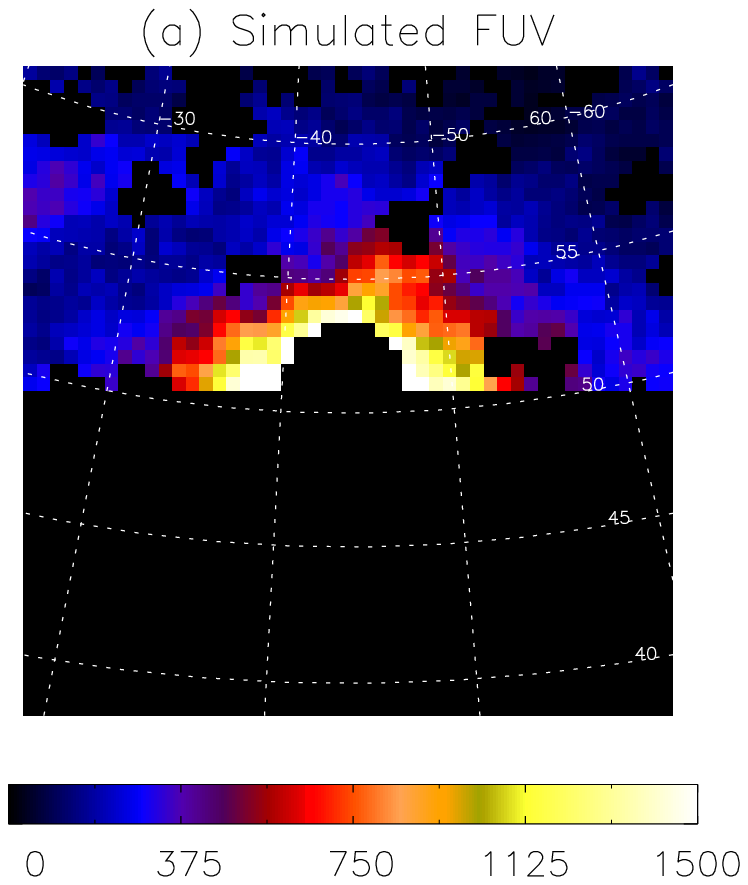}\hspace{0.4cm}
  \includegraphics[width=3.5cm]{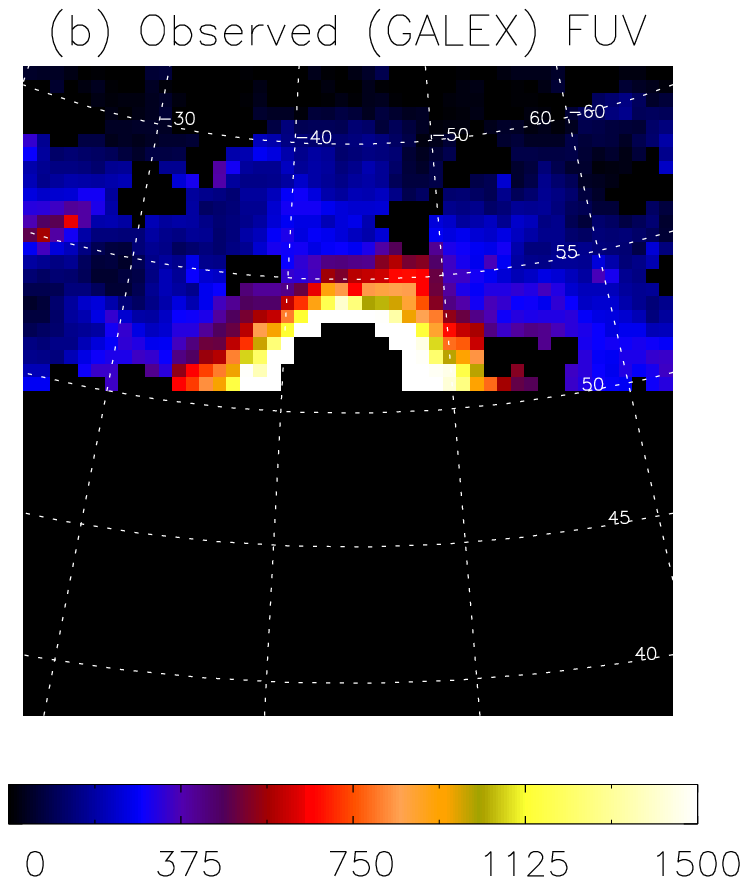}\\
  \vspace{0.8cm}
  \includegraphics[width=4.0cm]{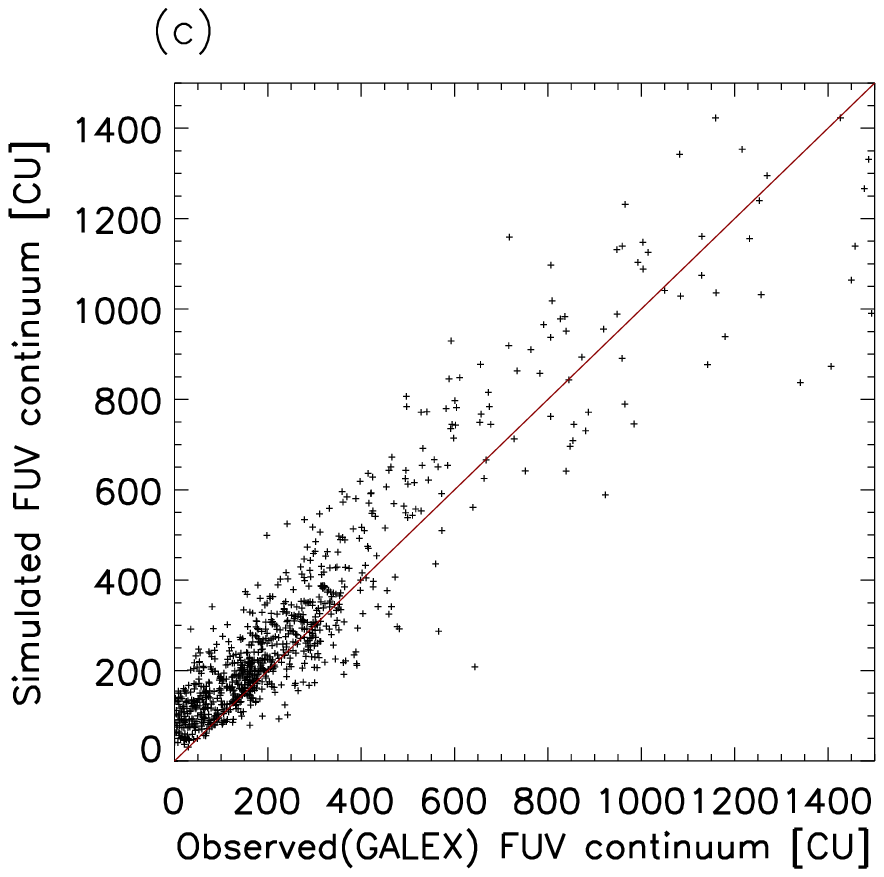}\\
  \end{center}
\caption{FUV continuum map of the northern region obtained from (a)
the best-fit dust scattering simulation, and (b) the GALEX
observation. A pixel-to-pixel correlation plot of the FUV intensity
obtained from (a) and (b) is shown in (c).\label{fig:north} }
\end{figure}

We also performed radiative transfer simulations for the northern
region, adopting the values of \textit{a} = 0.38 and \textit{g} =
0.46 and a single slab model. The distance to the front face and the
thickness of the dust slab were 70 pc and 30 pc, respectively. That
result implies that $\sim$33\% of the dust is located in front of
the central star and the approximate remaining 67 \% behind the
star. Figure \ref{fig:north} presents the resulting simulation map
together with the GALEX observation and compares the simulation
intensity with the GALEX intensity. As seen in the figure, the
simulation and observation are in good agreement. With $\sim$33\% of
the dust located in front of the central star, the \ion{H}{1} column
density of the northern region, obtained using the net mean
\textit{E}(\textit{B-V}) value of $\sim$0.015 after subtracting the
background of 0.015 and the relationship between
\textit{N}(\ion{H}{1}) and \textit{E}(\textit{B-V}) found in Figure
\ref{fig:gasdust}, is estimated to be $\sim$3 $\times$ 10$^{19}$
cm$^{-2}$, which is similar to that ($\sim$10$^{19}$ cm$^{-2}$)
obtained for the shell boundary of the LB \citep{cox87,lal03}.
Adopting a thickness of 10 pc, this gives a density of $\sim$0.9
cm$^{-3}$, which is consistent with those ($\sim$0.6 cm$^{-3}$ or
0.2 -- 0.6 cm$^{-3}$) estimated for the \ion{H}{2} region
\citep{rey85,par10}. While we used a single slab with a constant
density and thickness, it is remarkable that the results accord well
with those of other independent observations. We also tried two-slab
models, representing the LB and Loop I boundaries, but the two slabs
could not be separated effectively, perhaps because the density of
the LB is similar to that of the ambient medium and thus the present
models are not able to discriminate the shell of the LB from the
ambient medium.

\citet{sal08} analyzed the results of the FUV observations made by
the Far Ultraviolet Spectroscopic Explorer (FUSE) satellite towards
a single direction (\textit{l}, \textit{b}) = (277$^\circ$,
9$^\circ$) at the boundary of the interaction zone. By comparing the
\ion{O}{6} and \ion{C}{3} intensities of two other neighboring
directions across the boundary of the interaction zone, one just
outside the interaction zone and the other inside the interaction
zone, they were able to discriminate the contributions of Loop I
from those of the LB, by considering the fact that the interaction
zone with a neutral hydrogen column density of
\textit{N}(\ion{H}{1}) $\sim$ 4 $\times$ 10$^{20}$cm$^{-2}$ makes a
shadowing effect on the two FUV lines. The result shows that the
\ion{O}{6} emission comes mostly from the interface on the Loop I
side of the interaction zone while the \ion{C}{3} emission is
produced on the LB side of the interaction zone.

As the FIMS wavelength band includes the \ion{C}{4} doublet lines at
$\sim$1550 {\AA}, we conducted a similar study for regions A and B
in Figure \ref{fig:spectrum}(a), representing two distinct regions
with different extinctions: region B, being part of the interaction
zone, suffers more extinction for the emission originating from Loop
I than region A does. The average \textit{E}(\textit{B-V}) for
region A is $\sim$0.015 after subtracting the background extinction
of 0.015, which is translated into the optical depth of 0.11 at 1565
{\AA}, and the average value of \textit{E}(\textit{B-V}) for region
B is 0.075 after the background subtraction with a corresponding
optical depth of 0.54 at 1565 {\AA}. With the \ion{C}{4} intensities
I$_A$ = $\sim$5800$\pm$1100 LU and I$_B$ = $\sim$4300$\pm$ 1200 LU
determined in Section 2 for region A and B, respectively, we
obtained the intensities I$_{LB}$ = $\sim$ 1400$\pm$4000 LU and
I$_{LI}$ = $\sim$ 4900$\pm$5400 LU for the LB and Loop I,
respectively. The large error ranges are caused by the exponential
factors associated with extinctions when the observed uncertainties
of I$_{A}$ and I$_{B}$ are used to derived I$_{LB}$ and I$_{LI}$.
The result seems to imply that a larger portion of the \ion{C}{4}
emission comes from Loop I than from LB, although this is not
conclusive because of the large error range.


\section{Conclusions}

We have analyzed the datasets of the FUV observations performed by
GALEX and FIMS, together with the maps constructed in other
wavelengths, for the Spica Nebula and its neighborhood, which
includes the interaction zone of Loop I and the LB. The following
are the main findings of this study.

\begin{enumerate}
\item The diffuse FUV radiation observed in the northern
region above Spica was mainly due to the dust-scattering of
starlight originating from Spica, while the diffuse FUV radiation in
the southern region was attributed to scattering of photons
originating from background stars by dust in the interaction zone.
\item The FUV intensity showed a general correlation with the dust
color excess \textit{E}(\textit{B-V}).
\item The ratio of neutral hydrogen to dust was about $7.0\times10^{21}$
atoms cm$^{-2}$, which is slightly higher than previous studies
estimated for diffuse ISM.
\item Diffuse \ion{C}{4} emission was observed throughout the whole
region, including the interaction zone. By considering the shadowing
effect in the interaction zone, we found that a larger portion of
the \ion{C}{4} emission arises in the inner side of the shell
boundary Loop I than in the LB.
\item Molecular hydrogen fluorescence lines were also observed in the
interaction zone.
\item Based on the dust scattering simulation using a Monte Carlo
radiative transfer code, we estimated the optical parameters of dust
scattering as follows: \textit{a} = 0.38$\pm$0.06 and \textit{g} =
0.46$\pm$0.06, which are in agreement with previous radiative
transfer studies as well as the theoretical dust grain models of the
Milky Way.
\item We found that the interaction zone is located $\sim$ 70 pc away from the Sun, with a thickness of
$\sim$40 pc, covering the southern neighborhood of the Spica Nebula.
The central star Spica is likely to be located between the two
shells of LB and Loop I.
\end {enumerate}

\acknowledgments

FIMS/SPEAR is a joint project of KAIST and KASI (Korea) and UC
Berkeley (USA), funded by the Korea MOST and NASA grant NAG5-5355.
This research was supported by Basic Science Research Program
(2010-0023909) and National Space Laboratory Program (2008-2003226)
through the National Research Foundation of Korea (NRF) funded by
the Ministry of Education, Science and Technology. The Wisconsin
H-Alpha Mapper is funded by the National Science Foundataion.

\clearpage


\begin{thebibliography}{}
\bibitem[de Avillez \& Breitschwerdt(2012)]{avi12} de Avillez, M. A., \&
Breitschwerdt, D. 2012, \aap, 539, L1
\bibitem[Borkowski et al.(1990)]{bor90} Borkowski, K.
J., Balbus, S. A.,\& Fristrom, C. C. 1990, \apj, 355, 501
\bibitem[Bowyer(1991)]{bow91} Bowyer, S. 1991,
    \araa, 29, 59
\bibitem[Breitschwerdt et al.(2000)]{bre00} Breitschwerdt, D., Egger, R., \& Freyberg, M.
J. 2000, \aap, 361, 303
\bibitem[Breitschwerdt \& Avillez(2006)]{bre06} Breitschwerdt, D.,de Avillez,
M. 2006, \aap, 452, L1
\bibitem[Burhart \& Lazarian(2012)] {bur12} Burkhart,B., \&
Lazarian,A. 2012, \apjl, 755, L19
\bibitem[Burstein \& Heiles(1978)]{bur78} Burstein, D., \& Heiles,
C. 1978, \apj, 225, 40
\bibitem[Cardelli et al.(1989)]{car89} Cardelli, J.
A., Clayton, G. C., \& Mathis, J. S. 1989, \apj, 345, 245
\bibitem[Corradi et al.(2004)]{cor04} Corradi, W. J.
B., Franco, G. A. P., \& Knude, J. 2004, \mnras, 347, 1065
\bibitem[Cox \& Smith(1974)]{cox74} Cox, D. P., \& Smith, B. 1974, \apjl, 189, L105
\bibitem[Cox \& Reynolds(1987)]{cox87} Cox, D. P.,\& Reynolds, R.
J. 1987, \araa, 25, 303
\bibitem[Cox(1998)]{cox98} Cox, D. P. 1998, in IAU Colloq. 166, The
Local Bubble and Beyond, ed. D.Breitschwerdt, M. J. Freyberg, \& J.
Truemper (Lecture Notes in Physics, Vol. 506; New York: Springer),
121
\bibitem[Diplas \& Savage(1994b)]{dip94b} Diplas, A., \& Savage, B.
D.  1994a, \apjs, 93, 211
\bibitem[Diplas \& Savage(1994a)]{dip94a} Diplas, A., \& Savage, B.
D.  1994b, \apj, 427, 274
\bibitem[Dopita \& Sutherland(2001)] {dop01} Dopita, M. A., \&
Sutherland, R. S., Diffuse Matter in Universes, 2001, SpringerVerlag
\bibitem[Draine(2003)]{dra03} Draine, B. T. 2003,
    \apj, 598, 1017
\bibitem[Edelstein et al.(2006a)]{ede06a} Edelstein, J., Korpela, E. J., Adolfo, J., et al. 2006a,
    \apjl, 644, L159
\bibitem[Edelstein et al.(2006b)]{ede06b} Edelstein, J., Min, K. -W., Han, W., et al. 2006b,
    \apjl, 644, L153
\bibitem[Egger \& Aschenbach(1995)]{egg95} Egger, R., \& Aschenbach, B. 1995, \aap, 294, L25
\bibitem[Fejes(1974)]{fej95} Fejes.I. 1974, \aj, 79, 25
\bibitem[Ferlet et al.(1985)]{fer85} Ferlet R., Vidal-Major A.,\& Gry C. 1985, \apj, 298, 838
\bibitem[Finkbeiner(2003)]{fin03} Finkbeiner, D. P. 2003, \apjs, 146, 407
\bibitem[Fruscione et al.(1994)]{fru94} Fruscione, A., Hawkins, I., Jelinsky,
P., \& Wiercigroch, A. 1994, \apjs, 94, 127
\bibitem[G{\'o}rski et al.(2005)]{gor05} G{\'o}rski, K. ~M.,
Hivon, E., Banday, A.~J., et al. 2005, \apj, 622, 759
\bibitem[Haffner et al.(2003)]{haf03} Haffner, L. M.,
Reynolds, R. J., Tufte, S. L., et al. 2003, \apjs, 149, 405
\bibitem[Henyey \& Greenstein(1941)]{hen41} Henyey, L. G.,
    \& Greenstein, J. L. 1941, \apj, 93, 70
\bibitem[Henry(1991)]{hen91} Henry, R. C. 1991, \araa, 29, 89
\bibitem[Henry(2002)]{hen02} Henry, R. C. 2002, \apj, 570, 697
\bibitem[Hurwitz(1994)]{hur94} Hurwitz, M. 1994, \apj, 433, 149
\bibitem[Kalberla et al.(2005)]{kal05} Kalberla, P. M. W., Burton, W.
B.,Hartmann, D., et al. 2005, \aap, 440, 775
\bibitem[Jo et al.(2011)]{jo11} Jo, Y. -S., Min, K. -W., Seon, K. -I., et al. 2011,
    \apj, 738, 91
\bibitem[Jo et al.(2012)]{jo12} Jo, Y.-S., Min, K. -W., Lim, T. -H., Seon, K. -I., et al. 2012, \apj, 756, 38
\bibitem[Kim et al.(2007)]{kim07} Kim, I. -J., Min, K. -W., Seon, K. -I., et al. 2007, \apj, 665, L139
\bibitem[Kim et al.(2010a)]{kim10a} Kim, I. -J., Min, K, -W., Seon, K. -I., Han,
W., \& Edelstein, J. 2010a, \apj, 709, 823
\bibitem[Kim et al.(2010b)]{kim10b} Kim, I. -J., Seon, K. -I., Min, K. -W., et al.
2010b, \apj, 722, 388
\bibitem[Klessen(2000)]{kle00} Klessen, R.,S. 2000, \apj,535,869
\bibitem[Knapp \& Kerr(1974)]{kna74} Knapp, G. R., \& Kerr, F. J.
 1974, \aap, 35, 361
\bibitem[Knude (1978)]{knu78} Knude J. 1978, \aaps, 33, 347
\bibitem[Landini \& Monsignori Fossi(1990)]{lan90} Landini, M.,\&
Monsignori Fossi, B. C. 1990, \aaps, 82, 229
\bibitem[Lallement et al.(2003)]{lal03}
Lallement, R.,Welsh, B. Y., Vergely, J. L., Crifo, F., \& Sfeir, D.
 2003, \aap, 411, 447
\bibitem[Lee et al.(2006)]{lee06} Lee, D. -H., Yuk, I. -S., Jin, H., et al. 2006, \apj, 644, L181
\bibitem[Lee et al.(2008)]{lee08} Lee, D. -H., Seon, K. -I., Min, K. -W., et al. 2008,
\apj, 686, 1155
\bibitem[Lim et al.(2013)]{lim13} Lim, T. -H., Min, K. -W., \&
Seon, K.-I. 2013, \apj, 765, L107
\bibitem[Markwardt (2009)]{mar09} Markwardt, C. B. 2009, in ASP Conf.
Ser.411, Astronomical Data Analysis Software and Systems XVIII,ed.D.
A. Bohlender, D. Durand,\& P. Dowler (San Francisco, CA: ASP), 251
\bibitem[McKee \& Ostriker(1977)]{mck77} McKee, C. F.,\& Ostriker, J.
P. 1977, \apj, 218, 148
\bibitem[Morgan(1980)]{mor80} Morgan, D. H. 1980, \mnras, 190, 825
\bibitem[Morrissey et al.(2007)]{mor07} Morrissey, P.,Conrow, T.,\&
Barlow, T. A., et al. 2007, \apjs, 173, 682
\bibitem[Murthy \& Henry(2011)]{mur11} Murthy, J., \& Henry, R. C.
 2011, \apj, 734, 13
\bibitem[Paresce \& Jakobsen(1980)]{par80} Paresce,
F., \& Jakobsen, P. 1980, \nat, 288, 119
\bibitem[Park et al.(2007)]{par07} Park, J. -W., Min, K, -W., Seon, K. -I., et al. 2007,
 \apj, 665, L39
\bibitem[Park et al.(2010)]{par10} Park, J. -W., Min, K. -W.,
Seon, K. -I., Han, W., \& Edelstein, J. 2010, \apj, 719, 1964
\bibitem[Park et al.(2009)]{par09} Park, S. -J., Min, K. -W., Seon, K. -I., et
al. 2009, \apj, 700, 155
\bibitem[Park et al.(2012)]{psj12} Park, S. -J., Min, K. -W., Seon, K. -I., et
al. 2012, \apj, 754, 10
\bibitem[Reis \& Corradi(2008)]{rei08} Reis, W., \& Corradi, W. J.
B. 2008, \aap, 486, 471(RC2008)
\bibitem[Reynolds(1985)]{rey85} Reynolds, R. J. 1985, \aj,
90, 92
\bibitem[Reynolds(1990)]{rey90} Reynolds, R. J., 1990, in IAU Symp.
    Vol. 139, The galactic and extragalactic background radiation,
    ed. Bowyer, S., \& Leinert, C. (Dordrecht: Kluwer Academic), 157
\bibitem[Sallmen et al.(2008)]{sal08} Sallmen, S.
M., Korpela, E. J., \& Yamashita, H. 2008, \apj, 681, 1310
\bibitem[Savage \& Mathis (1979)]{sav79} Savage, B. D. \& Mathis,
J. S. 1979, \araa, 17, 73
\bibitem[Schlegel et al.(1998)]{sch98} Schlegel, D. J., Finkbeiner, D.
P., \& Davis, M. 1998, \apj, 500, 525
\bibitem[Seon(2003)]{seo03} Seon, K. -I. 2003, jkps, 43, 565
\bibitem[Seon et al.(2006)]{seo06} Seon, K. -I. 2006, \apjl, 644, L175
\bibitem[Seon et al.(2011a)]{seo11a} Seon, K. -I., Edelstein, J., Korpela, E. J., et al. 2011a,
    \apjs, 196, 15
\bibitem[Seon et al.(2011b)]{seo11b} Seon, K. -I., Witt, A. N., Kim, I. -J., et al. 2011b,
    \apj, 743, 188
\bibitem[Seon(2012a)]{seo12a} Seon, K. -I. 2012, \apjl, 761, L17

\bibitem[Seon \& Witt(2012b)]{seo12b} Seon, K. -I.,\& Witt, A. N. 2012,
    \apj, 758, 109
\bibitem[Seon(2013)]{seo13} Seon, K. -I. 2013, \apj, accepted (arXiv:1305.6144v1)
\bibitem[Shull \& van Steenberg(1985)]{shu85} Shull, J. M., \& van
Steenberg, M. E. 1985, \apj, 294, 599
\bibitem[Slavin(1989)]{sla89} Slavin, J. D. 1989, \apj, 346, 718
\bibitem[Sofue et al.(1974)]{sof74} Sofue,Y., Hamajima,
K.,\& Fujimoto, M. 1974, \pasj, 26, 399
\bibitem[Sternberg(1989)]{ste89} Sternberg, A., 1989, \apj,347,863
\bibitem[Sujatha et al.(2010)]{suj10} Sujatha, N. V., Murthy, J.,
Suresh, R., Henry, R. C., \& Bianchi, L. 2010, \apj, 723, 1549
\bibitem[Vazquez-Semadeni(1994)]{vaz94} Vazquez-Semadeni, E.
1994, \apj, 423, 681
\bibitem[Weingartner \& Draine(2001)]{wei01} Weingartner, J. ~C., \& Draine,
B. ~T. 2001, \apj, 548, 296
\bibitem[Welsh et al.(1994)]{wel94} Welsh, B.
Y., Craig, N., Vedder, P., \& Vallerga, J. V. 1994, \apj, 437, 638
\bibitem[Welsh \& Lallement(2005)]{wel05} Welsh, B. Y., \& Lallement,
R. 2005, \aap, 436, 615
\bibitem[Welsh et al.(2010)]{wel10} Welsh, B.
Y., Lallement, R., Vergely, J., \& Raimond, S. 2010, \aap, 510, 54
\bibitem[Willingale et al.
(2003)] {wil03} Willingale, R., Hands, A. D. P., Warwick, R. S.,
Snowden, S. L., \& Burrows, D. N. 2003, \mnras, 343, 995
\bibitem[Yoshioka \& Ikeuchi(1990)]{yos90} Yoshioka, S. \& Ikeuchi,
S. 1990, \apj, 360, 352
\bibitem[Yusef-Zadeh et al.(1984)]{yus84} Yusef-Zadeh, F., Morris, M.
    \& White, R. L. 1984, \apj, 278, 186
\bibitem[Zagury et al.(1998)]{zag98} Zagury, F., Jones, A.,
\& Boulanger, F.\ 1998, IAU Colloq. ~166: The Local Bubble and
Beyond, 506, 385
\end{thebibliography}
\end{document}